\documentclass[aps,prl,twocolumn,superscriptaddress,notitlepage,floatfix]{revtex4-2}
\usepackage{color}
\usepackage{amsmath}
\usepackage{amsthm}
\usepackage{amssymb}
\usepackage{graphicx}
\usepackage{float}

\usepackage{braket}
\usepackage{booktabs}

\usepackage{algorithm}  
\usepackage{algorithmicx}  
\usepackage{algpseudocode}

\usepackage[unicode=true,
 bookmarks=true,bookmarksnumbered=false,bookmarksopen=false,
 breaklinks=false,pdfborder={0 0 1},backref=false,colorlinks=true]
 {hyperref}
\hypersetup{
 linkcolor=magenta, urlcolor=blue, citecolor=blue, pdfstartview={FitH}, hyperfootnotes=false, unicode=true}

\makeatletter
\@ifundefined{textcolor}{}
{%
 \definecolor{BLACK}{gray}{0}
 \definecolor{WHITE}{gray}{1}
 \definecolor{RED}{rgb}{1,0,0}
 \definecolor{GREEN}{rgb}{0,1,0}
 \definecolor{BLUE}{rgb}{0,0,1}
 \definecolor{CYAN}{cmyk}{1,0,0,0}
 \definecolor{MAGENTA}{cmyk}{0,1,0,0}
 \definecolor{YELLOW}{cmyk}{0,0,1,0}
}

\usepackage{amsfonts,amsthm}\usepackage{tabularx}\usepackage{dcolumn}\usepackage{bm}\usepackage{graphicx}\usepackage{epstopdf}
\usepackage{times}
\usepackage[dvipsnames]{xcolor}

\usepackage{todonotes}
\usepackage{makecell}

\hypersetup{urlcolor=blue}

\makeatother

\begin{document}

\title{Prethermal Time-Crystalline Corner Modes}

\author{Si Jiang}
\affiliation{Center for Quantum Information, IIIS, Tsinghua University, Beijing 100084, China}

\author{Dong Yuan}
\affiliation{Center for Quantum Information, IIIS, Tsinghua University, Beijing 100084, China}

\author{Wenjie Jiang}
\affiliation{Center for Quantum Information, IIIS, Tsinghua University, Beijing 100084, China}

\author{Dong-Ling Deng}
\email{dldeng@tsinghua.edu.cn}
\affiliation{Center for Quantum Information, IIIS, Tsinghua University, Beijing 100084, China}
\affiliation{Hefei National Laboratory, Hefei 230088, China}
\affiliation{Shanghai Qi Zhi Institute, 41st Floor, AI Tower, No.~701 Yunjin Road, Xuhui District, Shanghai 200232, China}

\author{Francisco Machado}
\email{francisco.leal\_machado@cfa.harvard.edu}
\affiliation{ITAMP, Harvard-Smithsonian Center for Astrophysics, Cambridge, Massachusetts 02138, USA}
\affiliation{Department of Physics, Harvard University, Cambridge, Massachusetts 02138, USA}

\begin{abstract}
We demonstrate the existence of prethermal discrete time crystals whose sub-harmonic response is entirely localized to zero-dimensional corner modes.
Within the exponentially long prethermal regime, we show that the robustness of these corner modes arises from two related, yet distinct mechanisms: 
the presence of a higher-order symmetry-protected topological phase in the effective Hamiltonian, or the emergence of a dynamical constraint that prevents the decay of the corner mode.
While the first mechanism ensures the stability of the sub-harmonic response throughout the entirety of the prethermal regime, it is restricted to initial states in the ground state manifold of the effective Hamiltonian.
By contrast, the second mechanism enables the observation of the prethermal time-crystalline order for \emph{arbitrary initial states}, albeit with a time scale that is not only determined by the frequency of the drive, but also the relative energy scale across the system's sublattices.
We characterize these two mechanisms by simulating the dynamics of a periodically driven two-dimensional spin model, and discuss natural extensions of our model to all other dimensions.
\end{abstract}
\maketitle

Much like the spontaneous breaking of the spatial translation defines an ordinary crystal, the spontaneous breaking of time translation characterizes novel phases of matter termed \emph{time crystals}~\cite{Wilczek2012PRL, Shapere2012Classical}.
Since the initial proposals, much effort has been devoted to understanding the properties of time-crystalline order and under what conditions it can occur~\cite{Patrick2013Impossibility, Watanabe2015Absence, Vedika2016PRL, Else2016Floquet, Yao2017PRL,  Sacha2017Time, Khemani2019brief, Else2020Discrete, Zaletel2023Colloquium}.
Two broad questions motivate these explorations: 
what fundamentally determines the stability of a time crystal, and what is the nature of the long-lived subharmonic response?


When equations of motion are periodic in time, the corresponding discrete time crystal (DTC) is characterized by a robust and long-lived oscillation whose period is a multiple of the drive's~\cite{Vedika2016PRL, Else2016Floquet, Yao2017PRL,  Zhang2017Observation, Choi2017Observation, Mi2021Time, Frey2022Realization, Pal2018Temporal, Rovny2018Observation, Zeng2017Prethermal, Else2017Prethermal,  machado2020long, Kyprianidis2021Observation}.
In this setting, the stability of a DTC is predicated on the system's failure to equilibrate across all of the phase space,
a particularly challenging feat owing to interacting systems' natural ``tendency to be ergodic''~\cite{Prosen1998Time, Prosen1999Ergodic, Lazarides2014Equilibrium, Marin2015Universal, Alessio2014Long}.
One approach to overcome this challenge is to induce many-body localization (MBL)~\cite{Nandkishore2014Many-Body, Kjall2014Many-Body, Ponte2015Periodically, Abanin2019Colloquium, Ponte2015ManyBody, Lazarides2015Fate},
which offers the possibility for stabilizing a DTC to arbitrary late time~\cite{Vedika2016PRL, Else2016Floquet, Yao2017PRL, Zhang2017Observation, Choi2017Observation, Mi2021Time, Frey2022Realization, Pal2018Temporal, Rovny2018Observation}.
The dimensionality and disorder requirements of MBL have motivated the search for other settings where out-of-equilibrium phenomena can be stabilized, perhaps not to infinity, but rather parametrically long times.
The most well-known example is the prethermal discrete time crystal~(PDTC)~\cite{Zeng2017Prethermal, Else2017Prethermal, machado2020long, Kyprianidis2021Observation}, where the lifetime of the prethermal regime can be exponentially controlled by the frequency of the driving protocol~\cite{Abanin2015Exponentially, Kuwahara2016Floquet, abanin2017rigorous, Mori2018Floquet}, offering a large time window for studying the time-crystalline order. 
One big challenge is to understand under which settings this methodology can be extended.

\begin{figure}[!b]
    \centering
    \includegraphics[width=1\linewidth]{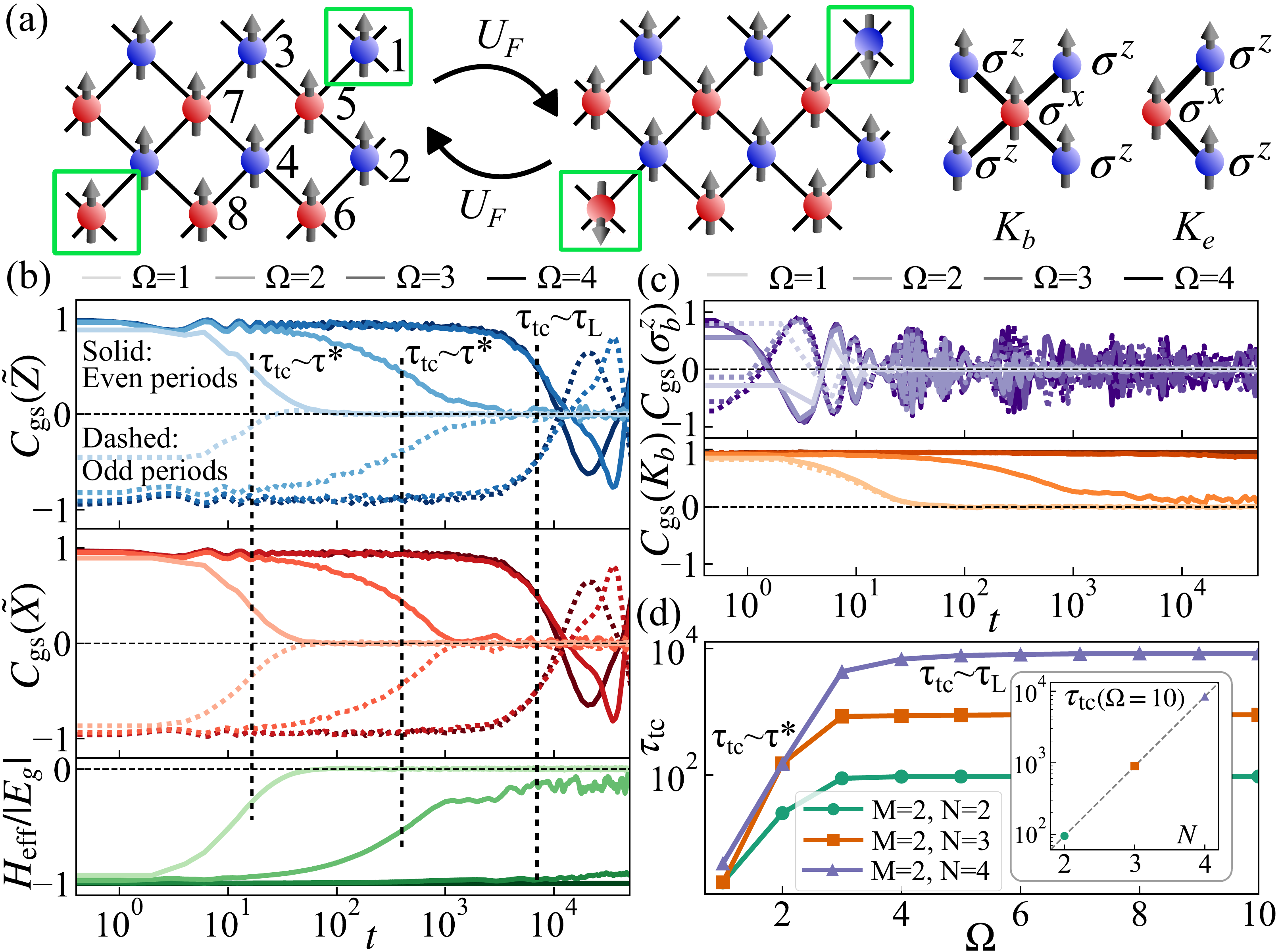}
    \caption{
    {\bf (a)} System composed of a checkerboard lattice with $1/2$-spins living on red and blue sublattices, and stabilizer operators $K_i$ in bulk and at edges. 
    Corners spins (green boxes) exhibit a subharmonic response under the Floquet unitary $U_F$.  
    {\bf (b)} Dynamics of the autocorrelation functions for corner mode operators $\tilde{Z}$, $\tilde{X}$, and energy density with different drive frequencies. The initial state is the ground state of $H_{\text{eff}}$. With low drive frequency ($\Omega =1,2$), the time-crystalline lifetime $\tau_{\text{tc}}$ matches the heating time $\tau^*$. With high drive frequency ($\Omega =3,4$), $\tau_{\text{tc}}$ is bounded by finite-size hybridization time $\tau_L$.
    {\bf (c)} Dynamics of the autocorrelation functions for $\sigma^z$ in the bulk which shows a beating behavior, and the stabilizer in the bulk which has the same periodicity as $H(t)$. 
    {\bf (d)} The time-crystalline lifetime $\tau_{\text{tc}}$ under varying drive frequencies and system sizes. The inset shows the exponential scaling of $\tau_{\text{tc}}$ with the distance between two corners under a high-frequency drive. These numerical results are obtained with $J= 1$, $V_{xx}=0.31$, $V_{zz}=0.15$, $\epsilon=0.05$, $h_x = 0.21, h_y = 0.17,h_z = 0.19$.}
    \label{fig:groundState}
\end{figure}

Another dimension for characterizing DTCs is the nature of observables that exhibit robust oscillations.
While the initial proposals involve spontaneous symmetry broken (SSB) systems where time-crystalline orders are observed in local observables throughout the whole system~\cite{Vedika2016PRL, Else2016Floquet, Yao2017PRL, Sacha2017Time, Zhang2017Observation, Choi2017Observation, Mi2021Time, Frey2022Realization, Pal2018Temporal, Rovny2018Observation, Zeng2017Prethermal, Else2017Prethermal,  machado2020long, Kyprianidis2021Observation}, it has been recently proposed and experimentally observed, that the same machinery can be ported to symmetry-protected topological (SPT) case or intrinsic topological order case, where different topological sectors are connected under the drive~\cite{zhang2022digital, wahl2023topologically, xiang2024longlived}. 
These examples demonstrate that time-crystalline responses can be restricted into subsystems.
However, since topological orders do not exist in finite temperature in two-dimensional (2D) and three-dimensional stabilizer codes~\cite{Yoshida2011Feasibility, Bravyi2009no-go}, one needs to prepare the systems' ground states to observe the robust subharmonic responses, which poses a stringent constraint on the initial states.





In this work, led by these two questions above, we discover a new kind of DTC phenomenon.
In particular, we construct a 2D Floquet system where the time translation symmetry is only broken in zero-dimensional subsystems (i.e. at its corners). 
The localized nature of these corners offers a new avenue for their late-time stabilization using the physics of prethermal strong zero modes (PSZMs)~\cite{Kemp2017Long, Else2017PrethermalZero, Fendley2016Strong, Parker2019Topologically, Kemp2020Symmetry, Yates2022Long, Mukherjee2024Emergent}. This removes any requirements on the energy density of the initial states~[Table.~\ref{tab:comparsion}].
Our results are threefold.
First, we show that under high-frequency driving, a higher-order symmetry-protected topological (HOSPT) phase~\cite{Isobe2015Theory, Huang2017Building, Song2017Topological, Song2017d-2, Schindler2018Frank, You2018Higher, Cifmmode2019Higher, Rasmussen2020Classification, Dubinkin2019Higher} emerges in the effective prethermal description. 
This phase gives rise to distinct gapless corner modes in the ground state manifold which exhibit long-lived subharmonic responses. Owing to the slow Floquet heating, the responses survive until an exponentially long heating time $\tau^*$ in the drive frequency.
Second, we observe that, while these corner modes are only topologically protected in the ground-state manifold and destroyed at finite temperature, the decay of corners into bulk is highly suppressed in a \emph{dimerized} version of our model.
This provides a distinct mechanism for the stabilization of long-lived time-crystalline order that extends to arbitrary initial states (i.e. finite temperature) and for an exponentially long time, controlled by the degree of dimerization.
Third, building upon the intuition in the 2D model, we discuss how to extend our results to arbitrary spatial dimensions and analyze how these time-crystalline dynamics can be realized on current superconducting quantum processors.

\textit{HOSPT-PDTC at zero temperature.---}We consider a  checkerboard lattice where spins-1/2 reside on two $N \times M$ sublattices [red and blue, Fig.\ref{fig:groundState}(a)] evolving under the following time-periodic Hamiltonian $H(t) = H(t+T)$:
\begin{align}
  \label{eq:evolution}
    H (t) = \begin{cases}
            \sum_{i} \left(\frac{\pi}{T}+\epsilon\right)\sigma_i^x & 0<t\leq \frac{T}{2} \\
            J\sum_{i\notin \text{corner}}  K_i+V(t) & \frac{T}{2}<t\leq T \\
       \end{cases}
\end{align}
Here, $\epsilon$ represents the imperfection of the $\pi$-pulse, and $K_i \equiv \sigma_i^x \prod_{\langle j i\rangle } \sigma_j^z$ is a multi-body stabilizer operator with $[K_a , K_b ] = 0$, where $\langle j i \rangle $ denotes the spins $j$ that are neighbors of spin $i$.
The stabilizer term $\propto \sum_{i\notin \text{corner}}  K_i$ is a cluster model that features a HOSPT phase in its ground state protected by the spin-flip symmetry $\prod_i \sigma^x_i$ on both red and blue sites, denoted by $\mathbb{Z}_2^{(b)}\times \mathbb{Z}_2^{(r)}$~\cite{You2018Higher, Dubinkin2019Higher, supplement}. 
$V(t)$ is a generic interacting perturbation that respects, at least,  $\mathbb{Z}_2^{(b)}$ or  $\mathbb{Z}_2^{(r)}$.
Throughout this work, we consider
\begin{align}
V(t) = J K_1+&h_x\sum_i{\sigma}_i^x+\sum_{i\in \text{red}}(h_y\sigma_i^y+h_z\sigma_i^z) \nonumber\\
&+V_{xx}\sum_{\langle i,j\rangle }\sigma_i^x\sigma_j^x+V_{zz} \sum P_i,
\end{align}
where $P_i=\prod_{j\in\square_i}\sigma^z_j$ are four-body plaquette terms in bulk or two-body terms at edges~\cite{supplement}.
We note that $V(t)$ explicitly breaks $\mathbb{Z}_2^{(r)}$ and thus the HOSPT phase is destroyed.

\begin{table}[t]
    \renewcommand{\arraystretch}{2}
    \caption{Different mechanisms for stabilizing a discrete time crystal in an isolated Floquet system. We explore time-crystalline order stabilized via prethermal higher-order symmetry-protected topological (HOSPT) phases at zero temperature (Fig.~\ref{fig:groundState}), and via prethermal almost strong zero modes at finite temperature (Fig.~\ref{fig:dimerization}).}
    \begin{tabular}{l||c|c|c}\hline \hline
    Mechanism    &  MBL  & \makecell{Prethermalization} & \makecell{Prethermal strong\\zero modes}   \\ \hline \hline
    \makecell{Origin of ergo-\\dicity breaking} & $\ell$-bits & \makecell{Emergent \\ symmetry} &  \makecell{Local obstruction \\ to equilibrium}\\ \hline
    Observable & \multicolumn{2}{c|}{\makecell{Local observables (SSB) \\  Edge modes [(HO)SPT]}} & \makecell{0D~Corner modes}\\
    \hline 
    Lifetime    & $\infty$ & $ e^{\Omega/\Omega_0}$ & $  e^{J/J_0} $\\ \hline
    Dimension       &  \makecell{Uncertain\\ for $d>1$} & \multicolumn{2}{c}{Arbitrary dimension}         \\ \hline
    Initial state      &  Any & \makecell{Below $T_c$ (SSB)\\ $T=0$ [(HO)SPT]}& Any \\ \hline\hline
    \end{tabular}
    \label{tab:comparsion}
\end{table}

When the driving frequency $\Omega = 2\pi/T$ is much larger than any local energy scale $\Omega_0$, the system’s dynamics under one period, $U_F = \mathcal{T}e^{-i\int_0^T H(t)dt}$, is well approximated by the evolution under an effective static Hamiltonian $H_\text{eff}$, followed by a perfect spin-flip $G = \prod_i \sigma_i^x = G_rG_b$.
Crucially, although $\mathbb Z_2^{(r)}$ is not a symmetry of the underlying dynamics, it is a symmetry of the effective Hamiltonian, $[H_{\text{eff}}, G_r]=0$, which is protected by the time-translation symmetry of the drive~\cite{Else2017Prethermal, machado2020long}. 
The ability of $H_{\text{eff}}$ to describe the system's dynamics is determined by the exponentially small error $O(e^{-\Omega/\Omega_0})$ in this approximation.
This error is thought of as the source of heating to $H_{\text{eff}}$ and, thus, determines an exponentially long prethermal window for $t<\tau^* \sim e^{\Omega/\Omega_0}$, where the system reach prethermal equilibrium with respect to $H_{\text{eff}}$:
\begin{align}
    2H_{\text{eff}} = J\sum_{i\notin \text{corner}}  K_i+&V_{zz} \sum_i P_i+V_{xx}\sum_{\langle i,j\rangle }\sigma_i^x\sigma_j^x \notag \\
    &+(h_x+\epsilon) \sum_i\sigma_i^x+O\left(\frac{\Omega_0}{\Omega}\right).\label{eq:H_eff}
\end{align}
When $V(t)=0$, $H_{\text{eff}}$ exhibits an exactly four-fold degeneracy of the ground-state manifold, corresponding to the presence of two gapless corner spin-1/2 modes. 
One of them is exactly localized at spin $1$ [Fig.~\ref{fig:groundState}(a)] and is characterized by two conjugate operators $\tilde{Z}=\sigma^z_1$ and $\tilde{X} = \sigma^x_1\sigma^z_5=K_1$. 
With non-zero perturbations $V(t)$, the corner mode becomes delocalized over a certain correlation length $\xi$.
Under the periodic drive, the time evolution under $H_{\text{eff}}$ preserves these corner modes, but the $G$ rotation flips both associated edge modes ($G\tilde{Z}G = -\tilde{Z}$ and $G\tilde{X}G = -\tilde{X}$). This induces a subharmonic response with the corner spin changing sign for each period, which is the defining feature of a PDTC.

Indeed, this is borne out by our numerical simulations.
Starting the system in the ground state of $H_{\mathrm{eff}}$, we compute the Floquet dynamics of Eq.~(\ref{eq:evolution}) with small $\epsilon$, $h_x$, $V_{xx}$ and $V_{zz}$ under different drive frequencies~[Fig.~\ref{fig:groundState}(b)].
For small frequencies ($\Omega = 1$, light curves), the rapid heating induces the fast decay of the autocorrelation functions $C_{\text{gs}}(\tilde{Z})=\text{Re}\langle \psi_{\text{gs}} | \tilde{Z}(t)\tilde{Z}(0) | \psi_{\text{gs}}\rangle$, $C_{\text{gs}}(\tilde{X})$ and energy density.
By contrast, at high frequencies ($\Omega = 4$, dark curves), the system does not thermalize after a very long time ($\tau^* > 10^4$); until then, $C_{\text{gs}}(\tilde{Z})$ and $C_{\text{gs}}(\tilde{X})$ exhibits a long-lived subharmonic response, consistent with conjugate corner modes whose responses lock to the subharmonic frequency $\omega/\Omega=1/2$.

A few remarks are in order.
First, the importance of the higher-order nature of the PDTC can be further emphasized by contrasting the dynamics of corners with either edges or bulk. 
Instead of a robust subharmonic oscillation, $C_{\text{gs}}(\sigma^z_{e/b})$ develops a strong beating behavior~[Fig.~\ref{fig:groundState}(c)], whose frequency is controlled by the precise parameters of the system (for more details in~\cite{supplement}). In addition, $C_{\text{gs}}(K_{e/b})$ features a periodicity matching that of $H(t)$.
Second, while in the thermodynamic limit, we expect the time-crystalline lifetime $\tau_{\text{tc}}$ to match the exponentially growing heating time scale $\tau^*$, in a finite-size system, the hybridization of the corner modes upper bounds $\tau_{\text{tc}}$.
This hybridization arises from the exponentially small overlap between the localized corner modes and induces an exponentially small gap $\Delta_{\text{FS}} \sim e^{-L/\xi}$ between different states of the ground state manifold, where $L$ is the distance between two corners.
This manifests itself in the dynamics of the autocorrelation function of the corner modes, which exhibits a long, coherent oscillation with a time scale $\tau_L\sim 1/\Delta_{\text{FS}}$.
Computing the frequency dependence of $\tau_{\text{tc}}$ demonstrates this phenomenon: after an initial exponential scaling $e^{\Omega/\Omega_0}$,  $\tau_{\text{tc}}$ plateaus to a size-dependent value that is exponentially controlled by the distance~[Fig.~\ref{fig:groundState}(d)].

\begin{figure}[t]
    \centering
    \includegraphics[width=1\linewidth]{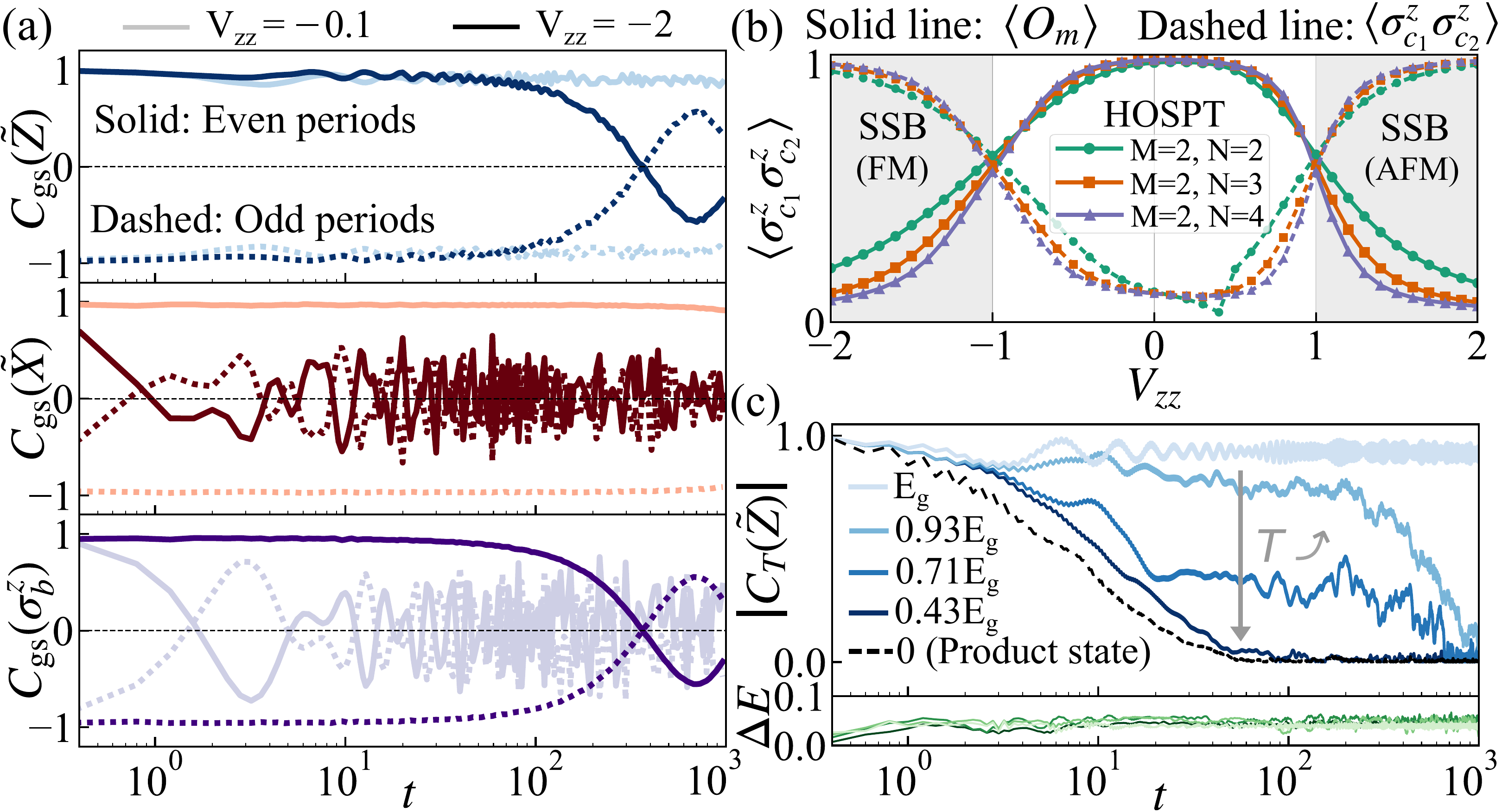}
    \caption{The melting of a HOSPT-PDTC. (a) When $|V_{zz}|$ is large, the time-crystalline behavior for $\tilde X$ is lost, and long-lived subharmonic responses occur for $\sigma^z$ at corners, edges, and in bulk, which indicates the system transits to an SSB-PDTC. (b) Characterizing the transition between HOSPT-PDTCs and SSB-PDTCs by membrane operator $\langle O_m\rangle$ and correlator at two blue corners $\langle \sigma_{c_1}^z\sigma_{c_2}^z\rangle$. The crossings indicate two transition points around $V_{zz}=\pm 1$. (c) The HOSPT-PDTC quickly melts when the initial state is not the ground state, even if the system has not thermalized yet.}
    \label{fig:3}
\end{figure}

\textit{Melting of a HOSPT-PDTC.---}The observed subharmonic response is robust to all kinds of perturbations in $V(t)$, as long as they preserve time-translation symmetry and the original $\mathbb{Z}_2^{(b)}$ symmetry. 
However, upon a large increase of either $h_x, V_{xx}$ or $V_{zz}$, $H_{\text{eff}}$ undergoes a quantum phase transition to a topologically trivial state, and the robustness of the corner modes is lost.
We analyze this effect in detail when tuning $V_{zz}$~[Fig.~\ref{fig:3}(a)], leaving the remaining perturbations to the Supplementary Materials~\cite{supplement}. 
When $|V_{zz}|$ is small (light curves), both conjugate operators $\tilde X$ and $\tilde Z$ exhibit robust time-crystalline behavior.
Upon increasing $V_{zz}$ (dark curve), the dynamics become different; while $C_{\text{gs}}(\tilde X)$ becomes fast decaying, $C_{\text{gs}}(\tilde Z)$ can still exhibit a long-lived subharmonic response.
In large $V_{zz}$ limit, $H_\mathrm{eff}$ supports ferromagnetic order (breaking $\mathbb{Z}_2$ symmetry) which stabilizes the observed sub-harmonic response.
We verify this picture by studying the response of the bulk and edge degrees of freedom and we observe that $C_{\text{gs}}(\sigma^z_{e/b})$ and $C_{\text{gs}}(\tilde Z)$ exhibit an equally long-lived behavior. 
At the same time, the ground state phase diagram of $H_{\text{eff}}$ as a function of $V_{zz}$ precisely exhibits a phase transition from a HOSPT phase to a ferromagnet at  $V^{c}_{zz}=-1$~[Fig.~\ref{fig:3}(b)], mirroring the transition from a HOSPT-PDTC to a SSB-PDTC~[Table.~\ref{tab:comparsion}].
Curiously, the locations of the transition are exactly pinned at $V_{zz}^c=\pm 1$ owing to a self duality of the model~\cite{supplement}.

One key ingredient in our discussions so far is that the system is initially prepared at zero temperature with respect to $H_{\text{eff}}$. Indeed, the topological protection of the gapless corner modes only exists within the ground state manifold~\cite{Yoshida2011Feasibility, Bravyi2009no-go}.
At finite temperature (i.e. higher energy density), thermal excitations from the bulk interact with these corner modes and lead to their decay.
As a result, even if the heating time scale of the system is very large, the time-crystalline behavior can exhibit a fast decay~[Fig.~\ref{fig:3}(c)]. 

\textit{Prethermal strong zero modes in a HOSPT-PDTC.---}While the decay of the corner modes into the bulk is a generic feature at non-zero temperature, this process is highly constrained by the symmetry properties of our model.
More specifically, the decay mainly arises from the energy exchange between stabilizer excitations at different sublattices.
This observation suggests a simple, yet powerful path towards the extension of the time-crystalline lifetime of the corner modes: By changing the relative strength of stabilizers between the two sublattices, i.e., by \emph{dimerizing} our model from $J$ to $J_r$ and $J_b$, the resonant process becomes suppressed and the corner modes become long-lived, regardless of the initial states.
This long lifetime signals the formation of PSZMs~\cite{Kemp2017Long, Else2017PrethermalZero, Fendley2016Strong, Parker2019Topologically, Kemp2020Symmetry, Yates2022Long, Mukherjee2024Emergent}.

The effectiveness of this dimerization strategy is best understood upon a dual transformation of $H_{\text{eff}}$~\cite{Kennedy1992Hidden, Doherty2009Identifying, You2018Subsystem, supplement}. The resulting dual model corresponds to two coupled plaquette Ising models, one for each sublattice:
\begin{align}
    2H_{\text{dual}} = J_r \sum_{i\in \text{red}} P_i+J_b\sum_{i\in\text{blue}} P_i +  V_{zz}\sum_{i\notin \text{corner}}K_i\notag\\+V_{xx}\sum_{\braket{i,j}}\sigma_i^x\sigma_j^x+(h_x+\epsilon)\sum_i\sigma_i^x + O\left(\frac{\Omega_0}{\Omega}\right)~.
\end{align}
With respect to the corner spins, the duality transformation maps them as follows: $\tilde Z\to \tilde Z_{\text{dual}}=\sigma_1^zG_r$ and $\tilde X\to \tilde X_{\text{dual}} = \sigma_5^z$, where $\sigma_{1,5}^z$ are at corners of the blue and red sublattices respectively.
Crucially, because $G_r$ commutes with the approximate Floquet evolution, the dynamics of $\tilde Z_{\text{dual}}$ are solely determined by the dynamics of $\sigma_1^z$~\footnote{Note that even if $G_r$ does not affect the dynamics of the $\tilde{Z}_\text{dual}$, it is still necessary to define the appropriate commutation relations between $\tilde{X}_\text{dual}$ and $\tilde{Z}_\text{dual}$.}. 
The robustness of the original corner operators is then mapped to the robustness of the magnetization of the two corner spins, which is easier to analyze because we only need to consider spin-flip processes.

The effect of the coupling between the two sublattices becomes transparent in this language.
Since a spin flip at corners either creates or destroys an odd number of plaquette excitations, when $J_r$ and $J_b$ are much larger than any other local energy scale $J_0$, this process becomes exponentially suppressed owing to the emergence conservation of excitation number~\cite{abanin2017rigorous}.
However, if we further add symmetry-preserved coupling between two sublattices (i.e. $V_{xx}$ terms), the process of flipping both corner spins at two sublattices can be resonant.
For example, $\sigma_1^x\sigma_5^x$ flips one $J_rP_i$ term and three $J_bP_i$ terms (one in bulk and two at edge). 
The resulting energy difference in the red sites is $2J_r$, and in the blue sites it is either $2J_b$ or $6J_b$ (depending on the particular spin configuration)~\footnote{Suppose $\sigma_1^z$ is $\downarrow$ and $\sigma_5^z$ is $\uparrow$, then the only configuration of $\sigma_5^z\sigma_6^z\sigma_7^z\sigma_8^z$ which resonates with $J_r=3J_b$ is $\uparrow\uparrow\uparrow\uparrow$. For all the other configurations, they will resonate with $J_r=J_b$.}. 
Therefore, when $J_r=J_b$ or $J_r=3J_b$, the process becomes resonant and the interaction strongly hybridizes the corners of different sublattices, which dominantly controls the corner modes' lifetime.
Crucially, when these resonant conditions are not met, the plaquette excitation number for each sublattice becomes approximately conserved, which will greatly enhance the corner spins' lifetime.

\begin{figure}[t]
    \centering
    \includegraphics[width=1\linewidth]{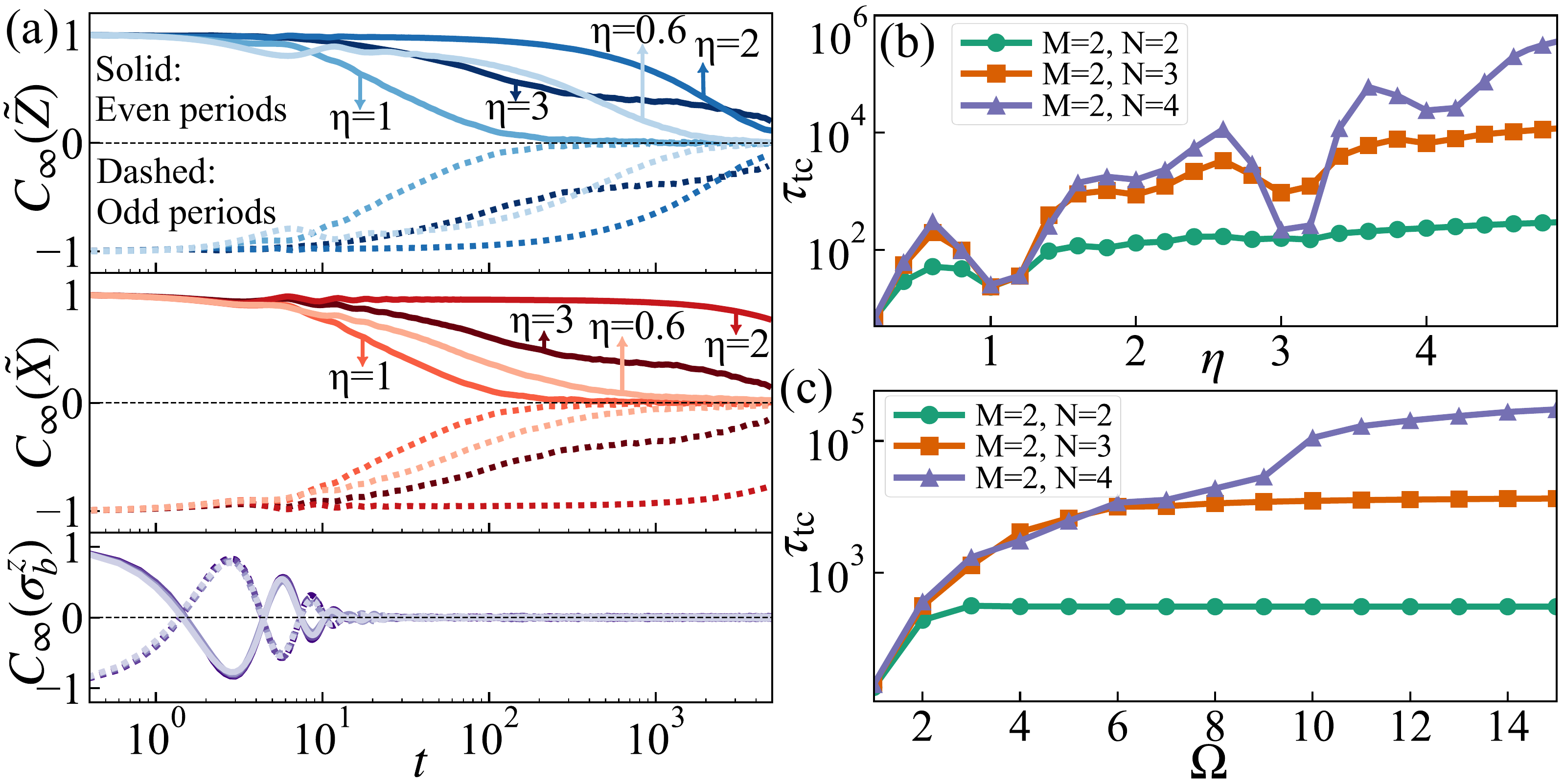}
    \caption{The prethermal strong zero modes in a HOSPT-PDTC at finite temperature. (a) Starting from a product state with all spins pointing towards $\hat z$, the lifetime for the subharmonic responses at corners is extended by several orders of magnitude through dimerizing relative energy scales between different sublattices, while the lifetime for edge and bulk degrees of freedom is not enhanced. (b) The time-crystalline lifetime $\tau_{\text{tc}}$ under different dimerization strengths and system sizes with large-frequency drive ($\Omega=20$). $\tau_{\text{tc}}$ first increases with $\eta$ and then saturates due to the finite system size. Two dips occur at $\eta=1,3$ due to the resonance. (c) $\tau_{\text{tc}}$ under different drive frequencies and system sizes with $\eta = 5.11$. These results are obtained with $J_b=1$, $V_{xx}=0.11$, $V_{zz}=0.05$,  $h_x = 0.11$. }
    \label{fig:dimerization}
\end{figure}

The above understanding is clearly demonstrated by our numerical investigations~[Fig.~\ref{fig:dimerization}(a)].
When studying the dynamics under a dimerization strength $\eta \equiv J_r/J_b \neq 1$, we observe a large enhancement of the lifetime of the corner's subharmonic response, even when starting from a product state in the middle of the spectrum~[Fig.~\ref{fig:dimerization}(a)]. 
In contrast, for spins at the edge or in the bulk, the lifetime remains almost unchanged as the excitations can move freely among these spins without changing the total excitation numbers.

It is important to contrast the nature of the protection of the corner modes between the non-dimerized (Fig.~\ref{fig:groundState}) and dimerized (Fig.~\ref{fig:dimerization}) cases.
In the former, the robustness of the corners arises from the presence of topologically protected gapless modes in the ground state manifold. As long as the initial state has a large overlap with the ground state of $H_{\text{eff}}$, the lifetime of the time-crystalline behavior is limited by the heating time of the system~$\tau_{\mathrm{tc}} \sim e^{\Omega / \Omega_0}$.
In contrast, the dimerization induces an \emph{energetic barrier} to the process that leads to the decay of the corners, and, thus, is not restricted to the ground state manifold.
In this case, the lifetime of the corners is limited by not only the heating time scale but also how well the dynamical constraint is satisfied, which (up to resonant conditions) can also be exponentially controlled~\cite{Kemp2017Long, Kemp2020Symmetry, Else2017PrethermalZero}.
The resulting lifetime is then upper bounded by $\tau_{\text{tc}}\sim\min\{e^{\Omega/\Omega_0}, e^{J_r/J_b}, e^{\min(J_r,J_b)/J_0}\}$.

In addition, the corner modes can be described by exact operators that commute, order by order, with $H_{\text{eff}}$. This further emphasizes the emergence of an approximately conserved quantity that constrains the system dynamics.
Starting with the zeroth-order term $\Psi_z^{(0)}=\tilde Z$, we consider perturbations of $h_x+\epsilon$, $V_{xx}$ and $V_{zz}$, which results in the first-order term:
\begin{equation}
    \Psi_z^{(1)} = \frac{h_x+\epsilon}{J_r} \Lambda_X + \frac{\eta^3}{(\eta^2-1)(\eta^2-9)}\frac{V_{xx}}{J_b}\Lambda_{XX},
\label{Eq:PsiExp}
\end{equation}
where $\Lambda_X$ and $\Lambda_{XX}$ consist of Pauli operators of spins around the corner~\footnote{$\Lambda_X = \sigma_1^x\sigma_2^z\sigma_3^z\sigma_4^z\sigma_5^x$, $\Lambda_{XX} = -
6\eta^{-3}\sigma_1^y\sigma_2^x\sigma_3^x\sigma_4^x\sigma_5^y\sigma_8^z+(7\eta^{-2}-1)\sigma_1^x\sigma_2^z\sigma_3^z\sigma_4^z+(3\eta^{-3}-\eta^{-1})\sigma_1^y\sigma_2^x\sigma_5^y\sigma_6^z+
(3\eta^{-3}-\eta^{-1})\sigma_1^y\sigma_3^x\sigma_5^y\sigma_7^z+2\eta^{-2}\sigma_1^x\sigma_2^y\sigma_3^y\sigma_4^z\sigma_6^z\sigma_7^z+2\eta^{-2}\sigma_1^x\sigma_2^z\sigma_3^y\sigma_4^y\sigma_6^z\sigma_8^z+2\eta^{-2}\sigma_1^x\sigma_2^y\sigma_3^z\sigma_4^y\sigma_7^z\sigma_8^z+(3\eta^{-3}-\eta^{-1})\sigma_1^y\sigma_4^x\sigma_5^y\sigma_6^z\sigma_7^z\sigma_8^z$.}, and the effect of $V_{zz}$ appears in higher-order terms. 
Curiously, $\Psi_z^{(1)}$ already encodes the information about the aforementioned resonances: $\Psi_z^{(1)}$ diverges when $\eta=1$ and $3$, which signals the breakdown of the perturbation analysis.
Performing our analysis to higher-order uncovers additional higher-order resonant processes for particular values of $\eta$ which further limit the corner mode lifetimes.
These higher-order resonances are visible in our analysis of the system's corner mode lifetimes [Fig.~\ref{fig:dimerization}(b)], with the emergence of dips in the lifetime at rational $\eta$.
Nevertheless, unless the interaction terms are tuned to these rational values, this perturbative approach predicts exponentially long lived corner modes, controlled by the strength of the preturbative terms \cite{Kemp2017Long, Abanin2015Exponentially, Abanin2017Effective}.
Indeed, for large $\eta$, $\tau_{\text{tc}}(\Omega)$ exhibits a similar behavior as the zero-temperature case~[Fig.~\ref{fig:dimerization}(c)], only being limited at large frequencies by system size hybridization of the corner modes.

We conclude our work by discussing future directions based on our results.
First, despite the multi-body nature of the stabilizer terms in our model, recent developments in superconducting qubit arrays have enabled the digital simulation of Floquet dynamics of complex correlated systems~\cite{Mi2021Time, Frey2022Realization, zhang2022digital, xiang2024longlived, Mi2022Noise}.
Building upon these developments, we propose an experimental protocol for the observation of these long-lived corner modes~\cite{supplement}. 
Second, while we have focused our attention on a 2D model, our results can be easily extended into higher spatial dimensions.
In this case, the $n$-dimensional stabilizer is defined on a hypercube, with a $\sigma^x$ at its center and a $\sigma^z$ operator for each of its corners.
The analogous stabilizers at $m$-dimensional edges will have $2^m$ $\sigma^z$ terms.
Crucially, all but the $m=0$ corner stabilizers will preserve the $\mathbb{Z}_2\times \mathbb{Z}_2$ symmetry.
As a result, the dynamically induced $\mathbb{Z}_2$ symmetry will cancel these corner stabilizers (in the high-frequency limit) and generate gapless corner modes.
In 1D, this construction corresponds to the ``ZXZ'' chain, which features an SPT phase for its ground state with two edge modes~\cite{Raussendorf2001OneWay, Chen2014Symmetry}.

Third, we note that the discussed PSZMs are not unique to HOSPT phases, but also exist in Kitaev chains, XYZ chains, and at interfaces between different systems and phases~\cite{Kemp2017Long, Fendley2016Strong, olund2023boundary}. It is then interesting if there exist novel features in time-crystalline orders of these systems.
Finally, the Floquet driving in our scenario generates only one $\mathbb{Z}_2$ symmetry, necessitating the underlying dynamics to respect another $\mathbb{Z}_2$ symmetry.
However, it is known that a quasi-periodic drive can generate multiple symmetries, offering a path for generating the fully required  $\mathbb Z_2\times \mathbb Z_2$ symmetry~\cite{Else2020Long, Friedman2022Topological, Dumitrescu2018Logarithmically, Dumitrescu2022Dynamical, he2024experimental}.
This comes at the cost of more complex Floquet heating dynamics. Understanding how the interplay of these effects affects the stability of the corner modes remains an open question.

\begin{acknowledgments}
We gratefully acknowledge the discussions with Jack Kemp about the prethermal strong zero modes, with Yong Xu and Kai Li about the higher-order symmetry-protected topological phases. The numerical simulations performed in this work are based on the dynamite~\cite{Gregory2023dynamite} and Cirq~\cite{Cirq2023cirq} package written in Python. S.J., D. Y., W. J., and D.-L.D. acknowledge support from the National Natural Science Foundation of China (Grant Nos.~T2225008,~12075128 and~123B2072), the Innovation Program for Quantum Science and Technology (Grant No.~2021ZD0302203), Tsinghua University Dushi Program, and Shanghai Qi Zhi Institute.
F.M. acknowledges support from the NSF through a grant for ITAMP at Harvard University.
\end{acknowledgments}

\bibliography{refs}
\bibliographystyle{apsrev4-1-title}

\clearpage
\newpage 
\onecolumngrid
\setcounter{section}{0}
\setcounter{equation}{0}
\setcounter{figure}{0}
\setcounter{table}{0}
\setcounter{page}{1}
\makeatletter
\renewcommand\thefigure{S\arabic{figure}}
\renewcommand\thetable{S\arabic{table}}
\renewcommand\theequation{S\arabic{equation}}

\begin{center} 
	{\large \bf Supplementary Materials: Prethermal Time-Crystalline Corner Modes}
\end{center} 

\setcounter{figure}{0}
\setcounter{table}{0}
\renewcommand\thefigure{S\arabic{figure}}
\renewcommand\thetable{S\arabic{table}}

\section{The system configurations and their corner modes}

\begin{figure}
    \centering
    \includegraphics[width=0.9\linewidth]{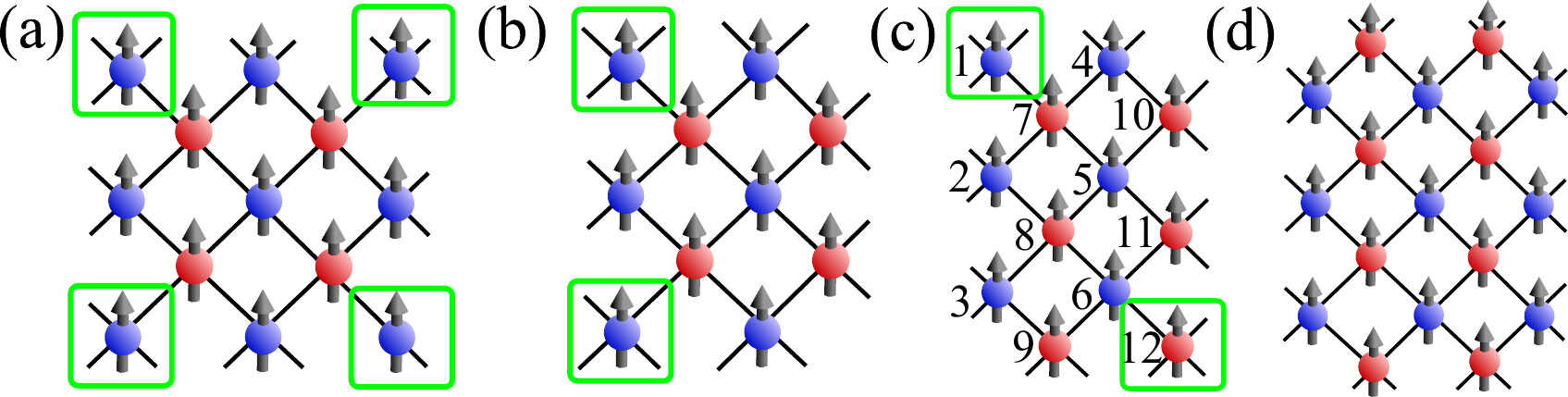}
    \caption{Different configurations for systems with rectangular shapes on the checkerboard lattice. The corners, which are highlighted by the green boxes, occur at the intersections of adjacent sides residing on the same sublattice type. }
    \label{fig:model_config}
\end{figure}

In the main text, we considered a system consisting of two sublattices of equal size ($N\times M$), such that each sublattice features a corner spin.
We argue that this requirement can be relaxed; by choosing different sizes for the two sublattices, the system has different number of corner spins, all of which exhibit the same features discussed in the main text.
Systems on the checkerboard lattice with rectangular shapes can be categorized into four groups, as shown in~Fig.~\ref{fig:model_config}(a-d): 
(a) all sides of the rectangle reside on the same sublattice; 
(b) there are three adjacent sides residing on the same sublattice; 
(c) there are two adjacent sides residing on the same sublattice, 
and (d) all adjacent sides reside on different sublattices. 
Corner sites occur when there are adjacent sides reside on the same sublattice. Their stabilizer operators break the $\mathbb{Z}_2\times \mathbb{Z}_2$ symmetry and will not occur in the prethermal effective Hamiltonian.

The degeneracy of the ground-state manifold can be computed by projecting the $\mathbb{Z}_2\times \mathbb{Z}_2$ symmetry onto the corners (which also leads to the operators describing the corner modes).
The symmetry generators $G_r=\prod_{i \in \text{red}}\sigma_i^x$, $G_b=\prod_{i \in \text{blue}}\sigma_i^x$ can be decomposed into the product of stabilizers and $\sigma_i^z$. 
Using the geometry shown in~Fig.~\ref{fig:model_config}(a) as an example:
\begin{equation}
    G_b = \prod_{i\in \text{blue}} K_i,\quad G_r = \prod_{i\in \text{red}} K_i \prod_{j\in \text{corner}}\sigma_j^z. 
\end{equation}
In the ground-state manifold of the fixed-point effective Hamiltonian (i.e. simple stabilizer Hamiltonian), these operators are further simplified by noting that the stabilizers at edges and in bulk take value $-1$. As a result, the action of the symmetry generators in the ground state manifold is given by:
\begin{equation}
    G_b \propto \prod_{i\in\text{corner}} K_i,\quad G_r \propto  \prod_{i\in \text{corner}}\sigma_i^z.
\end{equation}
Note that both generators are projected onto the corners, where $K_i$ and $\sigma_i^z$ anti-commute with one another.
Because the effective Hamiltonian commutes with both $G_r$ and $G_b$ (by the $\mathbb{Z}_2\times \mathbb{Z}_2$ symmetry), we immediately ensure that $\left[\prod_{i\in \text{corner}}\sigma_i^z, H_{\text{eff}}\right]=0$ and $[K_i, H_{\text{eff}}]=0$, which by locality of the Hamiltonian ensures that the corner modes $\tilde{Z}\equiv\sigma_i^z$ and $\tilde{X} \equiv K_i$ are integral of motions.
Together with the fact that $\tilde X$ and $\tilde Z$ are anti-commuted, we conclude that the degeneracy of the ground state manifold is greater than $1$, with the corner modes corresponding to the local operator that connects between the different ground states.

For the systems shown in~Fig.~\ref{fig:model_config}(b-c), similar process can also be done. For (b) we also have
\begin{equation}
    G_b \propto \prod_{i\in\text{corner}} K_i,\quad G_r \propto  \prod_{i\in \text{corner}}\sigma_i^z,
\end{equation}
and for (c),
\begin{equation}
    G_b \propto K_{\text{b,c}}\sigma_{\text{r,c}}^z ,\quad G_r \propto K_{\text{r,c}}\sigma_{\text{b,c}}^z
\end{equation}
where $K_{\text{r,c}}, \sigma_{\text{r,c}}^z\; (K_{\text{b,c}}, \sigma_{\text{b,c}}^z)$ are located at the red (blue) corner. However, for the system in (d), the symmetry generators become
\begin{equation}
    G_r = \prod_{i\in \text{red}} K_i \propto 1, \quad G_b=\prod_{i\in \text{blue}} K_i \propto 1,
\end{equation}
where the whole system has a unique ground state and no gapless corner modes exist.

\section{Analysing the system's dynamics}
As we argue in the main text, the system's dynamics are given by $U_F = e^{-iH_\text{eff}T}G+O(e^{-\Omega/\Omega_0})$ with the effective Hamiltonian being given by:
\begin{equation}
    2H_{\text{eff}} = J\sum_{i\notin \text{corner}}  K_i+V_{zz} \sum_i P_i+V_{xx}\sum_{\langle i,j\rangle }\sigma_i^x\sigma_j^x +(h_x+\epsilon) \sum_i\sigma_i^x+O\left(\frac{\Omega_0}{\Omega}\right).\label{eq:H_eff_supp}
\end{equation}
In the prethermal regime (while $U_F \approx e^{-iH_\text{eff}T}$ is a valid approximation), the autocorrelation function of the ground state after $n$ periods can be expressed, for any observable $O$, as 
\begin{equation}
    \begin{aligned}
    C_{\text{gs}}(O,nT)&=\Braket{\psi_0|O(nT)O(0)|\psi_0} \\
    &=\Braket{\psi_0|\prod^n\left(Ge^{iH_{\text{eff}}T}\right)O\prod^n\left(e^{-iH_{\text{eff}}T}G\right)O|\psi_0)} \\
    &= e^{iE_0nT}\Braket{\psi_0| G^n O G^n e^{-iH_{\text{eff}}nT} O|\psi_0},
\end{aligned}
\end{equation}
where $\ket{\psi_0}$ is the ground state of $H_{\text{eff}}$ and $E_0$ is the ground state energy. The last equation holds since $[G,H_{\text{eff}}]=0$. Since the operator $O$ we focus on either commutes or anti-commutes with $G$, we can define $G^nOG^n=f(O,n) O$ with
\begin{equation}
    f(O,n) = \begin{cases}
        1\quad & [O,G] = 0 \\
        (-1)^n & \{O,G\} = 0
    \end{cases},
\end{equation}
and the autocorrelation function is further simplified as 
\begin{equation}
    C_{\text{gs}}(O,nT) = f(O,n) e^{iE_0nT}\Braket{\psi_0|  O  e^{-iH_{\text{eff}}nT} O|\psi_0}. \label{eq:autocorrelation}
\end{equation}

The dynamics can be easily understood when the driving frequency is large ($\Omega_0/\Omega\to0$) and there are no perturbation terms ($\epsilon=h_x=V_{xx}=V_{zz}=0$). 
In this case, $H_{\text{eff}}=(J/2)\sum_{i\notin \text{corner}}K_i$ yields two exact gapless corner modes with four degenerated ground states. 
For $\tilde{Z}$ and $\tilde{X}$ which anti-commute with $G$ and commute with $H_{\text{eff}}$, we have $f(O,n) = (-1)^n$ and $O  e^{-iH_{\text{eff}}nT} O = O^2 e^{-iH_{\text{eff}}nT} = e^{-iH_{\text{eff}}nT}$, leading to
\begin{align}
    C_{\text{gs}}(\tilde Z,nT) =  C_{\text{gs}}(\tilde X,nT) = (-1)^n,
\end{align}
which are subharmonic responses with $\omega/\Omega = \frac{1}{2}$. For stabilizers $K_{e/b}$ at edges and in bulk which commute to both $G$ and $H_{\text{eff}}$, we have 
\begin{equation}
    C_{\text{gs}}(K_{e/b},nT) = 1.
\end{equation}
Situations change when we consider $\sigma_{e/b}^z$ which does not commute with $H_{\text{eff}}$. Concretely, at the edges and in bulk, $\sigma_i^z$ anti-commutes with $K_i$, but commutes with all $K_{j}$ with $j\neq i$. This leads to $e^{-iH_{\text{eff}}nT}\sigma_i^z=\sigma_i^ze^{-i(H_{\text{eff}}-JK_i)nT}$. Therefore, at edges and in bulk we have
\begin{equation}
    C_{\text{gs}}(\sigma_i, nT)=(-1)^n e^{-iJnT}.
\end{equation}
This is a rapid spin-flipping process modulated by a frequency determined by $J$, leading to the observed beating behavior, and thus {\bf not} a robust sub-harmonic response.

Now we consider the case when the perturbation terms are not zero. We first decompose $O\ket{\psi_0}$ into eigenstates of $H_{\text{eff}}$, which gives $O\ket{\psi_0} = \sum_i \alpha_i \ket{\psi_i}$. Then the autocorrelation function shown in~Eq.~\eqref{eq:autocorrelation} can be written as
\begin{equation}
\begin{aligned}
    C_{\text{gs}}(O,nT) &= f(O,n) e^{iE_0nT}\sum_{i,j} \alpha^*_j \alpha_i e^{-iE_inT} \braket{\psi_j|\psi_i}\\
    &= f(O,n) \sum_i |\alpha_i|^2 e^{-i(E_i-E_0)nT}
\end{aligned}.
\end{equation}
As long as the perturbations are small, $\tilde Z\ket{\psi_0}$ is very close to a system eigenstate $\ket{\psi_j}$ with $E_j-E_0\sim\Delta_{\text{FS}}$, where $\Delta_{\text{FS}}$ is the finite-size energy gap occurring at the ground-state manifold due to the perturbations. Therefore, we have 
\begin{equation}
    C_{\text{gs}}(\tilde Z,nT)\approx (-1)^ne^{-i\Delta_{\text{FS}} nT}.
\end{equation}
The same arguments can be carried out for $\tilde X\ket{\psi_0}$. This explains the very long, but finite system size lifetime $\tau_{L}$ observed in the main text.

\begin{figure}
    \centering
    \includegraphics[width=0.9\linewidth]{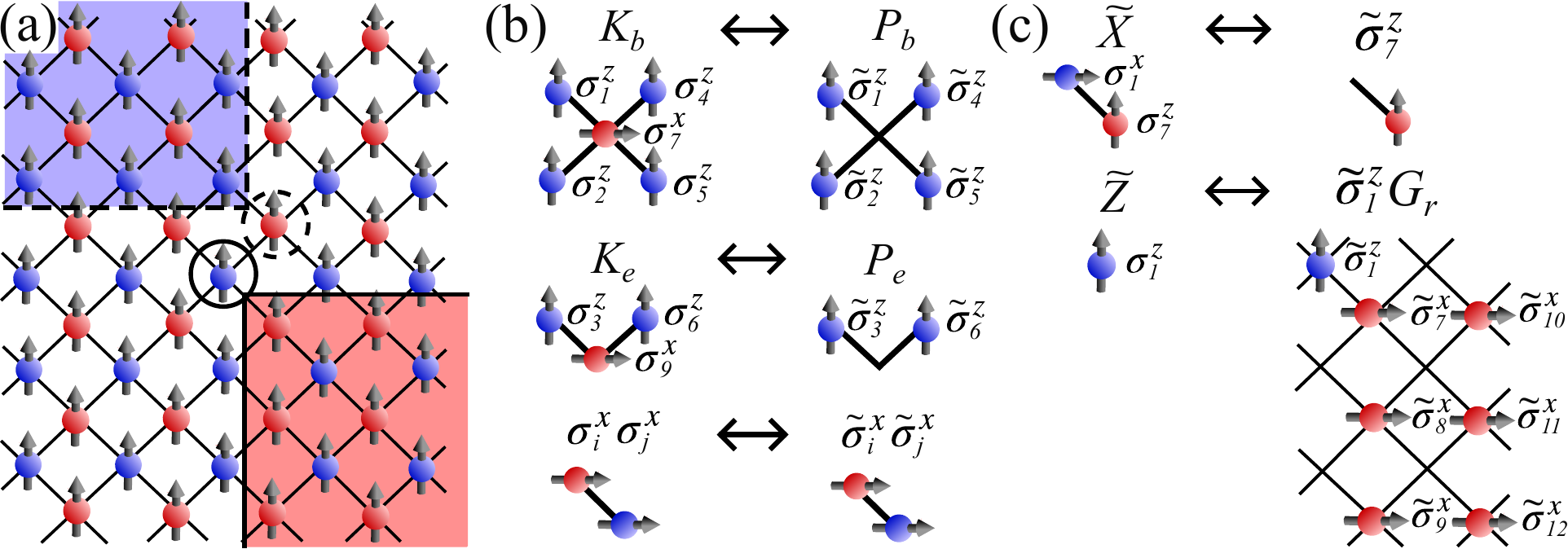}
    \caption{The duality transformation. (a) The $\sigma_i^z$ of a blue (red) spin is transformed into $\tilde\sigma_i^z$ times the product of $\tilde\sigma^x$ of the red (blue) spins in the red (blue) area $A_r$ ($A_b$). (b) The stabilizer operators in bulk (at edges) and the four-body (two-body) plaquette terms are transformed into each other. The $V_{xx} \sigma_i^x\sigma_j^x$ terms are transformed into themselves. (c) The corner operator $\tilde X$ is transformed into the $\tilde \sigma^z$ at the corner of the red sublattice, and $\tilde Z$ is transformed into the $\tilde \sigma^z$ at the corner of the blue sublattice times the emergent symmetry $G_r$. }
    \label{fig:dual}
\end{figure}

\section{The duality transformation}
In Eq.(3) of the main text, we show that the stabilizer operators are transformed to plaquette Ising terms on two sublattices in a dual picture. We now describe this duality in more detail. This duality is well-known in the one-dimensional case, which generalizes the idea in~\cite{Kennedy1992Hidden} that transforms a Haldane-type $\mathbb Z_2\times \mathbb Z_2$ chain into two Ising chains. The generalization of this duality into the two-dimensional case can be found in~\cite{Doherty2009Identifying}, which maps $\sigma^x$ to itself:
\begin{equation}
    \sigma_i^x \longleftrightarrow \tilde\sigma_i^x.
\end{equation}
While for $\sigma^z_i$, the mapping depends on the color of spin $i$:
\begin{equation}
    \sigma_i^z \longleftrightarrow \begin{cases}
        \tilde\sigma_i^z\prod_{j\in \text{red}, j\in A_r}\tilde\sigma_j^x,\quad & i\in \text{blue} \\
        \tilde\sigma_i^z\prod_{j\in \text{blue}, j\in A_b}\tilde\sigma_j^x,\quad & i\in \text{red} \\
    \end{cases},
\end{equation}
where $A_r$ contains all red spins located to the lower right of the spin $i$, and $A_b$ contains all blue spins located to the upper left of $i$, as shown in~Fig.~\ref{fig:dual}(a). 

With this duality transformation, we study how it acts on the prethermal effective Hamiltonian~Eq.~\eqref{eq:H_eff_supp}. One can verify that the stabilizer operators $K_i$ and the plaquette terms $P_i$ are mapped to each other, while $\sigma_i^x$ and $\sigma_i^x \sigma_j^x$ are mapped to themselves:
\begin{eqnarray}
    K_i = \sigma_i^x\prod_{\braket{ji}}\sigma_j^z\longleftrightarrow \prod_{\braket{ji}}\tilde\sigma_j^z = \tilde{P}_i, \\
    \sigma_i^x\longleftrightarrow \tilde{\sigma}_i^x, \quad  \sigma_i^x\sigma_j^x\longleftrightarrow \tilde{\sigma}_i^x \tilde{\sigma}_j^x.
\end{eqnarray}
We show concrete examples of transforming stabilizers in the bulk and at the edge in~Fig.~\ref{fig:dual}(b). Putting these results together, we obtain the effective Hamiltonian in the dual picture as
\begin{equation}
    2H_{\text{dual}} = V_{zz} \sum_{i\notin \text{corner}}  \tilde{K}_i+J \sum_i \tilde{P}_i+V_{xx}\sum_{\langle i,j\rangle }\tilde\sigma_i^x\tilde\sigma_j^x +(h_x+\epsilon) \sum_i\tilde\sigma_i^x+O\left(\frac{\Omega_0}{\Omega}\right).\label{eq:dual}
\end{equation}
$J\sum_i \tilde P_i$ is now the dominant term in the Hamiltonian, which can be considered as plaquette Ising models on two sublattices.
Note, that, by comparison with~Eq.~\eqref{eq:H_eff_supp}, $H_{\text{eff}}$ and $H_{\text{dual}}$ have the same form, which tells us that $H_{\text{eff}}$ is a self-dual model.
This directly gives that the phase transition between the plaquette Ising model and the HOSPT, which originates from the competition between $K_i$ and $P_i$, must occur at the self-dual point $V_{zz}/J =\pm 1$. 
This is in agreement with the observation of the HOSPT-PDTC to SSB-PDTC transition reported in the main text when $|V_{zz}|>|J|$.

We can also use this duality transformation to better understand the long lifetime of the corner modes. In the dual picture, the corner operators are mapped into
\begin{eqnarray}
    \tilde X = \sigma_1^x\sigma_7^z&\longleftrightarrow& \tilde \sigma_7^z \\
    \tilde Z = \sigma_1^z &\longleftrightarrow& \tilde \sigma_1^z \prod_{i\in \text{red}}\tilde\sigma_i^x = \tilde \sigma_1^z \tilde G_r,
\end{eqnarray}
which are shown in~Fig.~\ref{fig:dual}(c). Note that $\tilde\sigma_7^z$ and $\tilde \sigma_1^z$ are corners of the plaquette Ising models at the red and blue sublattices, which are protected by the domain wall conservation when $J$ dominates. In addition, $\tilde G_r$ becomes the emergent symmetry under the high-frequency drive. These together give the long-lived corner modes $\tilde Z$ and $\tilde X$.

\begin{figure}
    \centering
    \includegraphics[width=1\linewidth]{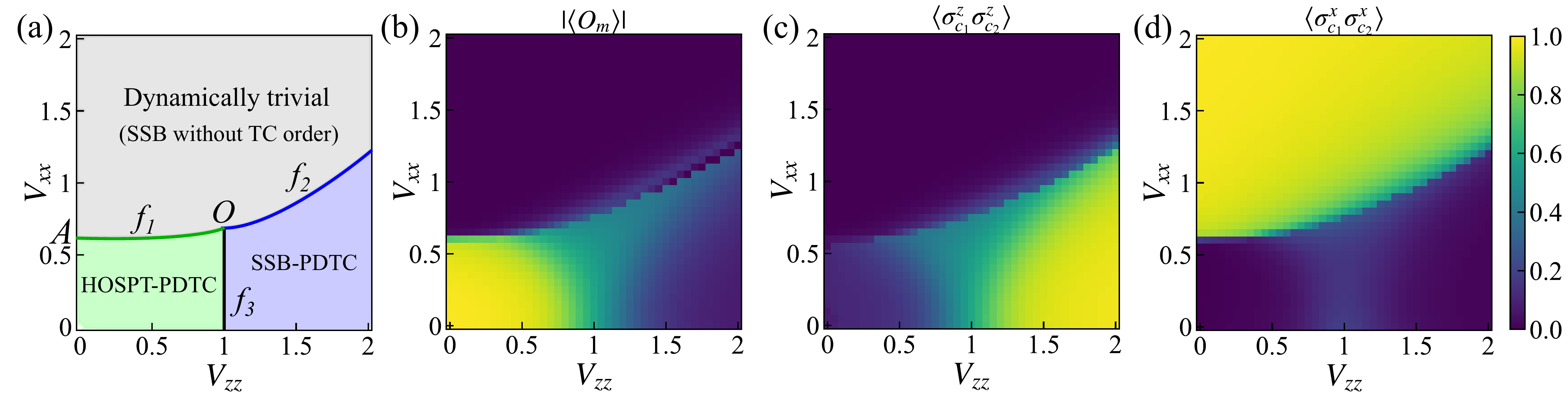}
    \caption{(a) The phase diagram as a function of perturbation strengths $V_{zz}$ and $V_{xx}$, where $f_1, f_2, f_3$ are the functions of phase boundaries, and $O$ is the intersection point of three phases. (b-d) Numerical calculations for the order parameters of the (b) HOSPT-PDTCs, (c) SSB-PDTCs, and (d) dynamically trivial phase. }
    \label{fig:phase}
\end{figure}
\section{The melting of HOSPT-PDTCs}
The long-lived subharmonic responses of the two conjugate corner operators $\tilde Z$ and $\tilde X$ originate from the HOSPT phase of $H_{\text{eff}}$~Eq.~\eqref{eq:H_eff_supp}. 
However, when any one of $h_x$, $V_{xx}$ and $V_{zz}$ is large enough, $H_{\text{eff}}$ undergoes a quantum phase transition and the sub-harmonic response is no longer robust---the PDTC melts.
In the previous section and the main text, we focused on the HOSPT-PDTC transition into an SSB-PDTC as $|V_{zz}|$ is increased.
We now discuss the remaining perturbations.

Intuitively, when $h_x$ is large, $H_{\text{eff}}$ develops a paramagnetic ground state that close to a product state with all spin pointing towards $-\hat x$.
In this case, both $\tilde Z$ and $\tilde X$ are destroyed and no time-crystalline order can be observed. 
In the dual picture, this corresponds to the destruction of the 2D plaquette Ising model by a large transverse field.

For the $V_{xx}\sum_{\braket{ij}}\sigma_i^x\sigma_j^x$ terms, when $V_{xx}$ is large enough, $H_{\text{eff}}$ develops a ferromagnetic ground state.
However, since the symmetry that is spontaneously broken $G = G_rG_b$ is exactly the same as the $\pi$-pulse $G$, we can not observe any time-crystalline order because there is no sub-harmonic response---the system becomes dynamically trivial.
We can also analyze the transition point in the dual picture, which is given by the transition point between a 2D Ising model with $XX$ terms and a 2D plaquette Ising model with $ZZZZ$ terms.

To summarize these results, we construct the full phase diagram as a function of $V_{zz}$ and $V_{xx}$~[Fig.~\ref{fig:phase}(a)]. 
In the green area with small $V_{zz}$ and $V_{xx}$, the system is a HOSPT-PDTC where the time-crystalline order can be observed on $\tilde Z$ and $\tilde X$.
In the blue area with small $V_{xx}$ but large $V_{zz}$, the system is an SSB-PDTC where the time-crystalline order can be observed at each $\sigma_i^z$ throughout the whole system.
In the grey area with large $V_{xx}$, the system enters a dynamically trivial phase, where the underlying phase of $H_{\text{eff}}$ is a ferromagnetic phase where $G$ is spontaneously broken.
By the self-duality, we know that the transition between HOSPT-PDTCs and SSB-PDTCs always occurs at $V_{zz}=1$, which means the phase boundary $f_3$ is a vertical line below the three-phase intersection point $O$. 

If $h_x=\epsilon=0$, we can further use the duality to obtain a relation between $f_1$ and $f_2$: We know that $(V_{zz}, f_2(V_{zz}))$ is the transition point between SSB-PDTC and the trivial phase under stabilizer strength $J=1$, and $(V_{zz}, f_1(V_{zz}))$ is the transition point between HOSPT-PDTC and trivial phase under plaquette Ising strength $V_{zz}$. 
Note that in the dual picture, the latter becomes the transition point $(1,f_1(V_{zz}))$ between SSB-PDTC and the trivial phase under stabilizer strength $\tilde J=V_{zz}$. Therefore, we have 
\begin{equation}
    \frac{f_1(V_{zz})}{V_{zz}}=f_2\left(\frac{1}{V_{zz}}\right).
\end{equation}
We can check that $f_1$ and $f_2$ indeed intersect at $O$ since $f_1(1)=f_2(1)$. Also, we can calculate the position of the transition point $A$ by
\begin{equation}
    \lim_{V_{zz}\to 0}f_1(V_{zz}) = \lim_{V_{zz}\to \infty}\frac{f_2(V_{zz})}{V_{zz}}.
\end{equation}

We numerically verify this phase diagram by calculating the order parameters $O_m=\prod_{i\in\text{red}, i\notin \text{corner}} K_i$ for HOSPT-PDTCs, $\braket{\sigma_{c_1}^z\sigma_{c_2}^z}$ for SSB-PDTCs, and $\braket{\sigma_{c_1}^x\sigma_{c_2}^x}$ for the dynamical trivial phase on a system with $M=N=3$. 
The results are shown in~Fig.~\ref{fig:phase}(b-d), where we can clearly identify the phase boundaries $f_1$, $f_2$, and $f_3$. Also note that $f_2(V_{zz})/V_{zz}\approx0.625$ at $V_{zz} = 2$ , which fits well with the transition point $A$ occurring at $V_{xx}\approx 0.6$.

\section{Experimental implementations and numerical simulations}
The superconducting transmon qubit arrays are a powerful platform for simulating quantum phases of matter~\cite{Mi2021Time, Frey2022Realization, zhang2022digital, xiang2024longlived, Mi2022Noise,deng2024highorder}. 
The HOSPT-PDTCs can be implemented on superconducting qubit arrays through a digital approach. To do this, we need to decompose the system dynamics $H_F(t)$ into sequences of experimentally implementable quantum gates.
The circuit $U_1$ for the $\pi$-pulse can be readily identified as 
\begin{equation}
    U_1=\prod_i R_{x,i}(\pi-\epsilon T).
\end{equation}
For the dynamics generated by stabilizer operators $J\sum_{i\notin \text{corner}}K_i$, we first note that the dynamics of one stabilizer can be implemented as follows:
\begin{equation}
\begin{aligned}
    U_{K_k} =& \exp\left(-i J_k\Delta t \bigotimes_{\braket{jk}}Z_j\otimes X_k \right ) \\
    =& \cos(-J_i\Delta t)\bigotimes_{\braket{jk}} I_j\otimes  I_k+i\sin(-J_k\Delta t) \bigotimes_{\braket{jk}}Z_j\otimes X_k \\
    =& \sum_{q_j\in\{0,1\}}\left[\cos(-J_k\Delta t)\bigotimes_{\braket{jk}}\ket{q_j}\bra{q_j}\otimes  I_k+i\sin(-J_k\Delta t) \bigotimes_{\braket{jk}}(-1)^{q_j}\ket{q_j}\bra{q_j}\otimes X_k \right]\\
    =& \sum_{q_j\in\{0,1\}}\left[\bigotimes_{\braket{jk}}\ket{q_j}\bra{q_j}\otimes  Z_k^{\sum q_j}[\cos(-J_k\Delta t)I_k+i\sin(-J_k\Delta t)\otimes X_k ]Z^{\sum q_j}_k\right]\\
    =& \left[\prod_{\braket{jk}} \text{CZ}_{j,k} \right]R_{x,k}(2J_k\Delta t)  \left[\prod_{\braket{jk}} \text{CZ}_{j,k} \right] 
\end{aligned}\label{eq:PauliDecom}
\end{equation}
where the stabilizer operator is applied at $k$ and $\braket{jk}$ denotes the spins $j$ that are neighbors of spin $k$. 
We note that all the two-qubit gates involved in this circuit are between nearest-neighboring qubits, which suit well the qubit connection geometry for existing programmable superconducting processors. 
By adjusting the rotation angle of the single qubit $R_x$ gate, we effectively implement the evolution with arbitrary stabilizer strength and driving frequencies. 

\begin{figure}[t]
    \centering
    \includegraphics[width=0.9\linewidth]{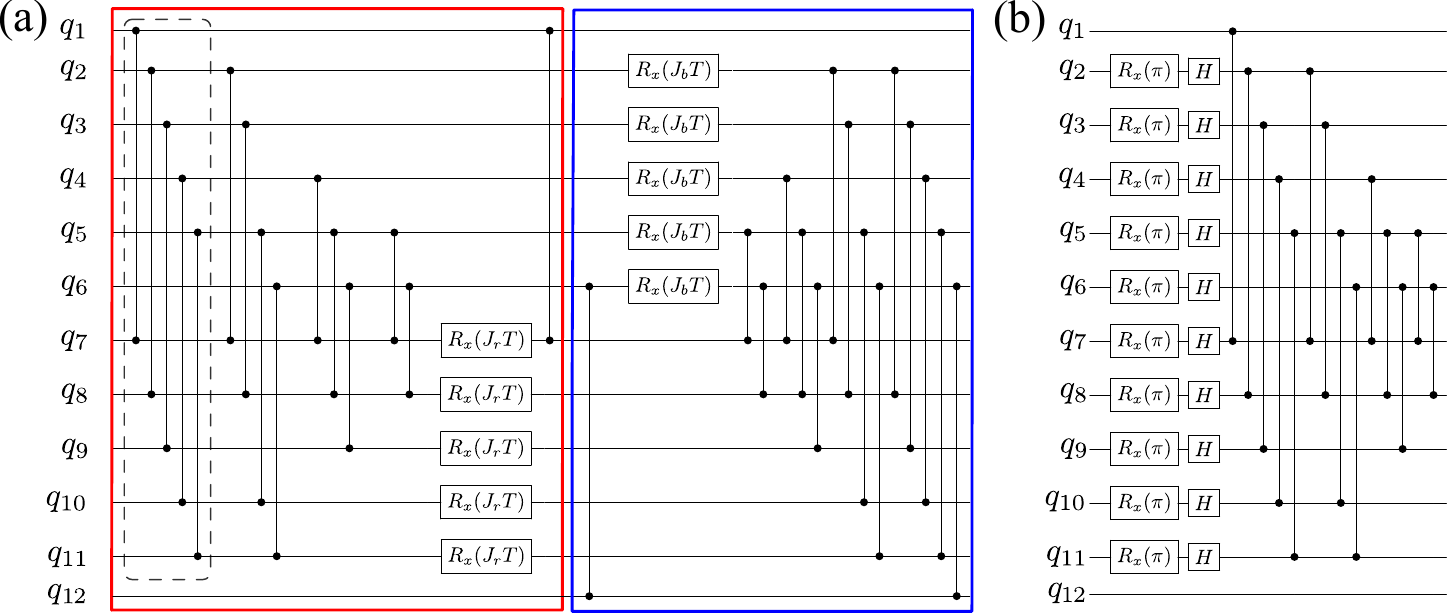}
    \caption{The quantum circuits for (a) simulating the dynamics generated by stabilizer operators, and (b) preparing the system ground state of $J\sum_{i\notin \text{corner}}K_i$ with $M=2, N=3$.
    The qubit labels are the same as the ones shown in~Fig.~\ref{fig:model_config}(c).
    All two-qubit CZ gates involve near-neighboring qubits.}
    \label{fig:circuits}
\end{figure}

To implement $J\sum_{i\notin \text{corner}}K_i$, note that all stabilizer operators commute, we only need to place $U_{K_k}$ together and arrange them with minimized circuit depth. 
This optimization involves first implementing stabilizer operators on qubits located on one sublattice, followed by implementation on another. 
The optimized circuit $U_2$ for a $M=2, N=3$ system is shown in~Fig.~\ref{fig:circuits}(a). 
In this setup, qubits $q_1\sim q_6$ reside on the blue sublattice, and qubits $q_7\sim q_{12}$ reside on the red sublattice [labeled in~Fig.~\ref{fig:model_config}(c)]. 
The dashed box represents a two-qubit gate block, where CZ gates can be executed on distinct sites simultaneously during experiments. 
The gates in the red (blue) box implement the stabilizers at qubits residing on the red (blue) sublattice. 
Except for those connected with corners, the CZ gates between red and box boxes are mutually canceled.
Notably, all two-qubit CZ gates are applied on adjacent qubits. It is worth emphasizing that this circuit is scalable, maintaining the same circuit depth regardless of the system size, thereby facilitating simulations for large systems on NISQ devices.

We note that the stabilizer operator at the blue corner $JK_1$ in $V(t)$ can also be implemented exactly by the method discussed above.
Other perturbation terms in $V(t)$ are difficult to simulate exactly since, in general, they do not commute with $J\sum_{i\notin \text{corner}}K_i$. 
Nevertheless, note that the effect of perturbation terms is to break the integrability and break the $\mathbb{Z}_2\times \mathbb{Z}_2$ symmetry of $J\sum_{i\notin \text{corner}}K_i$. 
This can also be done by adding other circuits after $U_2$. For example,  we can use the following circuits to approximate perturbations by single-body Pauli terms and two-body $XX$ terms.
\begin{equation}
    U_{\text{single}} = \prod_{i} R_{x,i}(h_xT)\prod_{i\in \text{red}} R_{y,i}(h_y T)\prod_{i\in \text{red}} R_{z,i}(h_z T),
\end{equation}
\begin{equation}
    U_{xx} = \prod_{k\in \text{blue}} \prod_{\braket{jk}} \text{CX}_{j,k}R_{x,k}(2V_{xx}\Delta t)\text{CX}_{j,k}.
\end{equation}

Besides simulation of the system evolution, we also need to prepare the ground state $\ket{\psi_g}$ of $J\sum_{i\notin \text{corner}}K_i$ as the initial state to see the dynamics of HOSPT-PDTCs at zero temperature. 
To do this, we follow the idea in~\cite{Satzinger2021Realizing}, where the authors use it to prepare the ground state of the toric code model. 
In our case, note that the ground state satisfies $\braket{\psi_g|K_i|\psi_g} = -1$ for all $i\notin \text{corners}$, we can write it as 
\begin{align}
    \ket{\psi_g} &\propto \prod_{i\notin \text{corners}} (I-K_i)\ket{0}^{\otimes N_{\text{qubits}}}.
\end{align}
To implement $I-K_i$, we can again use the scheme shown in~Eq.~\eqref{eq:PauliDecom}, which gives
\begin{equation}
    \begin{aligned}
    I-K_i &= \bigotimes_{\braket{ji}} I\otimes  I-\bigotimes_{\braket{ji}}Z\otimes X \\
    &=\sum_{q_j\in\{0,1\}}\left[\bigotimes_{\braket{ji}}\ket{q_j}\bra{q_j}\otimes  I-(-1)^{\sum q_j} X\right]\\
    &=\sum_{q_j\in\{0,1\}}\left[\bigotimes_{\braket{ji}}\ket{q_j}\bra{q_j}\otimes  Z^{\sum q_j}(I-X)Z^{\sum q_j}\right]\\
    &=\left[\prod_{\braket{ji}} \text{CZ}_{j,i} \right](I-X)\left[\prod_{\braket{ji}} \text{CZ}_{j,i} \right].
\end{aligned}
\end{equation}
Since we start with $\ket{0}^{\otimes N_{\text{qubits}}}$, the ground state preparation can be further simplified as 
\begin{equation}
    \begin{aligned}
    \ket{\psi_g} &\propto \prod_{\substack{i\in \text{red} \\ i\notin \text{corners}}} (I-K_i)\prod_{\substack{i\in \text{blue} \\ i\notin \text{corners}}} (I-K_i)\ket{0}^{\otimes N_{\text{qubits}}} \\
    &\propto \prod_{\substack{i\in \text{red} \\ i\notin \text{corners}}} (I-K_i)\ket{0}_{\text{b,c}} \bigotimes_{\substack{i\in \text{blue} \\ i\notin \text{corners}}  } \ket{-} \bigotimes_{\substack{i\in \text{red} }  } \ket{0} \\
    & \propto \prod_{\substack{i\in \text{red} \\ i\notin \text{corners}}} \left[\prod_{\braket{ji}} \text{CZ}_{j,i} \right] \ket{0}_{\text{b,c}} \ket{0}_{\text{r,c}} \bigotimes_{\substack{i\in \text{blue} \\ i\notin \text{corners}}  } \ket{-} \bigotimes_{\substack{i\in \text{red} \\ i\notin \text{corners}} } \ket{-} \\
    & \propto \prod_{\substack{i\in \text{red} \\ i\notin \text{corners}}} \left[\prod_{\braket{ji}} \text{CZ}_{j,i} \right] \prod_{i\notin \text{corners}} H_iX_i \ket{0}^{\otimes N_{\text{qubits}}}.
\end{aligned}
\end{equation}
The circuit $U_{\text{gs}}$ for preparing the ground state is shown in~Fig.~\ref{fig:circuits}(b).

\begin{figure}
    \centering
    \includegraphics[width=0.9\linewidth]{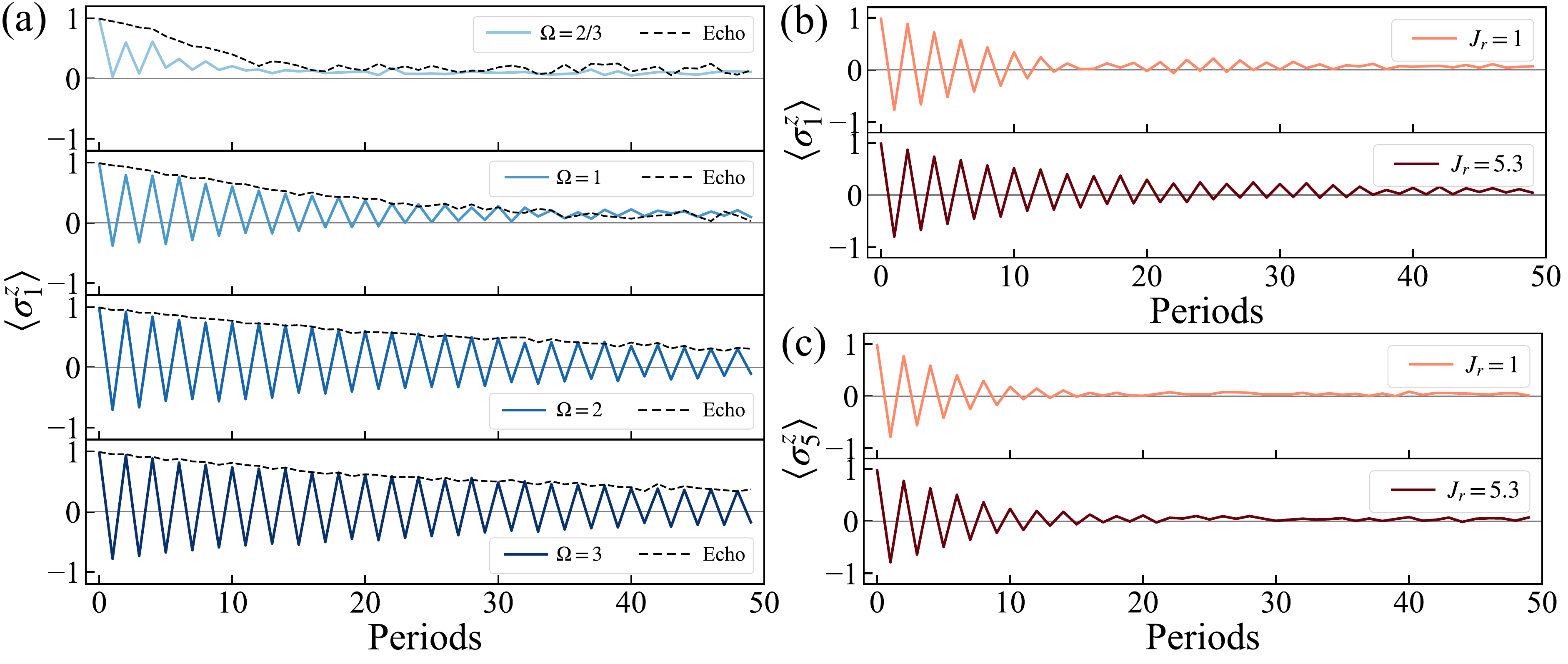}
    \caption{Noisy simulations for dynamics of a HOSPT-PDTC. (a) The dynamics of corner spin magnetization at zero temperature. With increasing driving frequency, the dynamics show a long-lived subharmonic response with an envelope saturated with the echo dynamics envelope. (b-c) The dynamics of (b) corner spin magnetization and (c) bulk spin magnetization at finite temperatures. With increasing dimerization strength, only the corner spin lifetime of the subharmonic response is extended.}
    \label{fig:exp_simu}
\end{figure}

To show that time-crystalline orders of HOSPT-PDTCs can be observed on NISQ devices, we carry out numerical simulations with a realistic noise model. 
This is done with the help of the quantum virtual machine provided in the Cirq package in Python~\cite{Cirq2023cirq}. We run the virtual machine named ``Weber,'' which simulates running quantum circuits with noise data that mimic Google's Sycamore processor. 
We first simulate the HOSPT-PDTCs at zero temperature, which is done by first applying $U_{\text{gs}}$ and then repeatedly applying $U_F=U_{\text{single}} U_2U_1$.
We simulate different driving frequencies by changing $T$ in $U_2$ and $U_{\text{single}}$. After each period, the magnetization at the blue corner is measured and gives $\braket{\sigma_1^z(nT)}=\braket{U_\text{gs}^\dagger (U_F^\dagger)^n \sigma_1^z(U_F)^nU_\text{gs}}$. The simulation results are shown in~Fig.~\ref{fig:exp_simu}(a). 
We identify that the magnetization quickly decays to zero under low-frequency drive. In contrast, under high-frequency drive, the magnetization shows a long-lived subharmonic response. 
To verify that the quick decay in the low-frequency case is due to the thermalization of the system's intrinsic dynamics but not the experimental noise, we also simulate the echo circuit $U_{\text{echo}} = U_{\text{gs}}^\dagger (U_F^\dagger)^n (U_F)^n U_{\text{gs}}$ and measure $\sqrt{\braket{U_{\text{echo}}^\dagger \sigma_1^z U_{\text{echo}}}}$. 
This value measures the effect of the experimental errors and noise~\cite{Mi2021Time}. 
By increasing the driving frequencies, the envelope of $\braket{\sigma_1^z(nT)}$ gradually saturates to the echo envelope. This indicates that under high-frequency drive, the system thermalizes very slowly by intrinsic dynamics, and the lifetime of the prethermal regime is extended.

At finite temperatures, the effectiveness of dimerization is also observed. Beginning with $\ket{0}^{\otimes N_{\text{qubits}}}$, we simulate different dimerization strength by changing $J_r$ in $U_2$. 
We repeatedly apply $U_F = U_{xx}U_2U_1$ and measure the spin magnetization $\sigma_1^z$ at the corner and  $\sigma_5^z$ in bulk. The simulation results are shown in~Fig.~\ref{fig:exp_simu}(b-c). We can identify that with larger dimerization strength ($J_r = 5.31$), the lifetime of the subharmonic response at the corner is extended, while the one in bulk remains almost unchanged.
\end{document}